\documentstyle[tighten,aps]{revtex}

\twocolumn

\newcommand{\be}{\begin{equation}}
\newcommand{\ee}{\end{equation}}
\newcommand{\ba}{\begin{eqnarray}}
\newcommand{\ea}{\end{eqnarray}}

\newcommand{\non}{\nonumber}

\def\vec#1{{\mbox{\boldmath$#1$}}}

\begin{document}

\draft

\title{
\[ \vspace{-2cm} \]
\noindent\hfill\hbox{\rm physics/0105108} \vskip 10pt
Ionization Potential of the Helium Atom}

\author{ Vladimir Korobov\thanks{
e-mail:  korobov@thsun1.jinr.su}}
\address{ Joint Institute for Nuclear Research,
Dubna, 141980, Russia}
\author{ Alexander Yelkhovsky\thanks{
e-mail:  yelkhovsky@inp.nsk.su}}
\address{ Budker Institute of Nuclear Physics,
Novosibirsk, 630090, Russia} \maketitle

\begin{abstract}
Ground state ionization potential of the He$^4$ atom is evaluated
to be $5\;945\;204\;221\,(42)\,{\rm MHz}$. Along with lower order
contributions, this result includes all effects of the relative
orders $\alpha^4$, $\alpha^3 m_e/m_\alpha$ and $\alpha^5\ln^2
\alpha$.
\end{abstract}


In contrast to the theoretical description of electromagnetically
bound two-body systems like hydrogen, positronium or muonium,
where considerable progress is achieved (for the recent reviews
see, e.g., \cite{hydrogenII}), high precision calculations in more
complex atoms are elaborated to a lesser degree. The central
problem with an extension of the methods developed for the
two-body problem to few-electron atoms is that those methods are
usually strongly rely on the solution of the Schr\"odinger
equation for a single particle in the Coulomb field. Having a
simple analytic form, this solution is a perfect reference point
for the calculation of various observables as power series in the
fine structure constant $\alpha$ using the quantum-mechanical
perturbation theory. In higher orders of this perturbation theory,
where the nonrelativistic approximation usually breaks down, the
explicit form of the nonrelativistic solution facilitates the
extraction of the ultraviolet divergences. These divergences are
canceled by matching to their finite counterparts calculated in
the fully relativistic framework of the quantum electrodynamics.

Although the Schr\"odinger equation for, e.g., three particles
bound by the Coulomb potentials can be solved numerically with
very high accuracy \cite{mp}, the lack of an analytic solution
makes the problem of the divergences cancellation more involved as
compared to the two-body case. This problem was recently analyzed
in Ref. \cite{heops} using singlet states of the helium atom as an
example. Employing the nonrelativistic quantum electrodynamics
\cite{cl} regularized dimensionally, it is demonstrated in
\cite{heops} how all the divergences arising in the
quantum-mechanical perturbation theory can be extracted and
canceled at the \emph{operator} level, without recourse to an
explicit form of the helium wave function. For the first time the
${\cal O}(\alpha^4)$ correction to singlet $S$ levels of the
He$^4$ atom is represented as a sum of apparently finite average
values of the regularization-independent operators.

In this Letter we present the most precise evaluation of the
helium ground state energy. Expressed in terms of the ionization
potential (the difference between ground state energies of the
singly charged ion and of the atom), our result reads:
\begin{equation}
\nu_{\rm th}(1^1 S) = 5\;945\;204\;221\,(42)\,{\rm MHz}.
\label{result}
\end{equation}
Along with the nonrelativistic energy, this result includes all
${\cal O}(\alpha^2)$, ${\cal O} (\alpha^3)$, ${\cal O}(\alpha^4)$
and ${\cal O}(\alpha^5 \ln^2 \alpha)$ relativistic and radiative
corrections. We take into account the finite nucleus-to-electron
mass ratio $M\equiv  m_\alpha/m_e = 7\,294.299\,508 (16)$
\cite{cuu} exactly in the nonrelativistic and ${\cal O}(\alpha^2)$
contributions, include the first $(\sim 1/M)$ recoil correction
into the ${\cal O}(\alpha^3)$ contribution and neglect the nucleus
recoil in higher orders. The effect of a finite nucleus charge
radius $R_{\rm N} = 1.673(1)\,{\rm fm}$ \cite{br} is included into
the helium ground state energy as
\begin{equation}
\delta_{\rm chr} E =
      \frac{2\pi Z \alpha}{3} R_{\rm N}^2
          \left\langle \delta(\vec{r}_1) +
          \delta(\vec{r}_2) \right\rangle.
\label{chrad}
\end{equation}
Here $Z = 2$ is the nucleus charge (in units of the proton one),
while $\vec{r}_1$ and $\vec{r}_2$ denote the positions of the
electrons with respect to the nucleus. The angle brackets in
(\ref{chrad}) and below denote the average value over the
nonrelativistic ground state. In (\ref{result}) we take one half
of the ${\cal O}(\alpha^5 \ln^2 \alpha)$ correction as an estimate
of the uncertainty due to higher orders. Our result agrees with
the previous theoretical estimate
\begin{equation}
\nu^{\rm DM}_{\rm th}(1^1 S) = 5\;945\;204\;226\, (91)\,{\rm MHz},
\label{DM}
\end{equation}
obtained in \cite{dm} and including the ${\cal O}(\alpha^4)$ and
${\cal O}(\alpha^3/M)$ effects only partially\footnote{The
uncertainty in (\ref{DM}) exceeds that in (\ref{result}) because
only part of the ${\cal O}(\alpha^4)$ corrections has been
included into the calculation that lead to Eq.(\ref{DM}).}.

In the remaining part of this Letter we briefly describe details
of our calculation. Ground state energy of the helium atom is
calculated as power series in the fine structure constant
$\alpha$. Leading ($\,\sim\! 1$) contribution\footnote{Unless
otherwise specified, we use the atomic units $e=\hbar=m_e=1$ and
$c = 1/\alpha$ throughout this paper. In particular, the unit of
energy is $2\hbox{Ry}\equiv m_e c^2 \alpha^2$.}, the Schr\"odinger
energy $E$, and the corresponding wave function $\psi$ are found
as a solution of the variational problem
\begin{equation}
E = \min_\psi \frac{\left\langle \psi | H | \psi \right\rangle}
                   {\left\langle \psi | \psi \right\rangle},
\end{equation}
for the helium atom Hamiltonian taken in the nonrelativistic
approximation,
\begin{equation}
H = \frac{p_1^2 + p_2^2}{2} + \frac{P^2}{2M}
    - \frac{Z}{r_1} - \frac{Z}{r_2} + \frac{1}{r}.
\label{nrH}
\end{equation}
Here $r_{1,2} = |\vec{r}_{1,2}|$ and $r = |\vec{r}_1 -
\vec{r}_2|$; $\vec{p}_{1,2}$ are the momenta of the electrons and
$\vec{P}= -\vec{p}_1 -\vec{p}_2$ is the momentum of the nucleus.

To construct the variational wave function we use the simplest
form of the basis,
\begin{equation}
\psi_n = \exp ( - k_1^n r_1 - k_2^n r_2 - k_3^n r), \;\;\; n =
1,\ldots, N. \label{basis}
\end{equation}
The complex exponents $k_a^n$ are chosen in a quasi-random manner
from a rectangular area on the complex plane, for example,
\begin{equation}
{\rm Re}\; k_a^n = K_a^{\rm min} +
        \left\lfloor \frac{n(n+1)}{2} \sqrt{p_a} \right\rfloor
        \left(K_a^{\rm max} - K_a^{\rm min} \right),
\end{equation}
where $\lfloor x\rfloor$ denotes a fractional part of $x$, $p_a$
is some prime number, while $\left(K_a^{\rm min}, K_a^{\rm max}
\right)$ is a variational interval. Imaginary parts of the
parameters are generated in a similar way. We use both real and
imaginary parts of $\psi_n$ to form a set of real basis functions.
In particular, the ground state wave function $\psi$  is a linear
combination of $2N$ basis functions ${\rm Re}\; \psi_n,$ ${\rm
Im}\; \psi_n$, $n=1,\ldots, N$, symmetric over the interchange of
the electrons positions, $r_1 \leftrightarrow r_2$.

Variational expansion in the basis (\ref{basis}) was shown in
\cite{mp} to be very effective. It yields the best available
nonrelativistic energies for many atomic and molecular systems and
in particular for the ground state of the helium atom. Simplicity
of the basis (\ref{basis}) allows us to evaluate analytically
matrix elements of all the operators that appear in the
calculation. By a proper differentiation and/or integration of the
basic integral,
\ba
\int d^3\vec{r}_1 \int d^3\vec{r}_2 \frac{
     \exp(-k_1 r_1 - k_2 r_2 - k_3 r )}{r_1 r_2 r}&& \non \\
    = \frac{16 \pi^2}{(k_1+k_2)(k_2+k_3)(k_3+k_1)}&&,
\ea
with respect to $k_1$, $k_2$ and $k_3$ we express the matrix
element of any operator involved in our calculation in terms of
rational functions of $k$'s, their logarithms and dilogarithms.

For the zeroth order approximation a wave function built within a
set of $2N=1200$ basis functions has been used that yields the
nonrelativistic energy
\begin{equation}
E = - 2.903\>304\>557\>727\>940\>23(1). \label{E}
\end{equation}
Here and below we cite the uncertainty of the numerical results
due to finiteness of the basis set. The uncertainties due to
incomplete knowledge of the physical constants are included into
the final result for the ionization potential (see Table). High
accuracy of (\ref{E}) is not redundant since the calculation of
rather singular matrix elements of higher order corrections
requires very accurate variational wave function.

First relativistic correction to the nonrelativistic value
(\ref{E}) is the average of the Breit Hamiltonian (see, e.g.,
\cite{BSbook}) over $\psi$:
\ba
\delta^{(2)} E &=& \alpha^2
             \left\langle - \frac{ p_1^4 + p_2^4 }{ 8 }
                 - \frac{ P^4 }{ 8M^3 }
                 + \pi Z \frac{\delta(\vec{r}_1)
                 + \delta(\vec{r}_2)}{ 2 }
                 \right.\non \\
            && +\; \pi \delta(\vec{r})
             - \frac{1}{ 2 }
      \left( \vec{p}_1 \frac{1}{r} \vec{p}_2 +
      (\vec{p}_1\vec{n})\frac{1}{r}
      (\vec{n}\vec{p}_2) \right)
\non \\
  &&\left.
      + \frac{Z}{ 2 M } \left( \vec{p}_1 \frac{1}{r_1}
      \vec{P}
+ (\vec{p}_1\vec{n}_1)\frac{1}{r_1}(\vec{n}_1\vec{P}) +
(1\!\to\!2) \right) \right\rangle
               \non \\
               &=& -1.952\,050\,77(1)\, \alpha^2.
\label{breitcorr}
\ea
Here $\vec{n} = \vec{r}/r$ and $\vec{n}_{1,2} =
\vec{r}_{1,2}/r_{1,2}$. To simplify the presentation, we
explicitly take into consideration that the spin of the nucleus
and the total spin of electrons are both equal to zero. In
particular, we replace the product of the electron spin operators
$\vec{s}_1 \vec{s}_2$ by its eigenvalue in the singlet state,
$-3/4$.

Order $\alpha^3$ and $\alpha^3/M$ corrections to the energy can be
represented as follows (see \cite{heops} and references therein):
\ba
\delta^{(3)}E &=& \alpha^3
  \left\{\frac{4Z}{3}
     \left[-2\ln\alpha-\beta+\frac{19}{30}
     \right]
     \left\langle \delta(\vec{r}_1)+\delta(\vec{r}_2)
     \right\rangle
  \right. \non \\
&& +\;\left(\!
    \frac{14}{3}\ln\alpha+\frac{164}{15}\!\right)
       \left\langle \delta(\vec{r}) \right\rangle
    +\frac{7}{3\pi}
       \left\langle \frac{\ln r\! + \gamma}{r^2}
                 i\vec{n} 
                 \vec{p}
    \right\rangle
    \non \\
&& +\;\frac{2Z^2}{3M}
    \left(-\ln\alpha-4\beta+\frac{31}{3}\right)
    \left\langle \delta(\vec{r}_1) + \delta(\vec{r}_2)
    \right\rangle \non \\
&& \left. +\; \frac{7Z^2}{3\pi M}
            \left\langle \frac{ \ln{r_1} + \gamma }{ r_1^2 }
            i\vec{n}_1 
            \vec{p}_1
       + (1\to 2)
                 \right\rangle\right\}
\non \\
&=&\; 57.270\,34(2)\, \alpha^3. \label{e3}
\ea
Here $\gamma = 0.5772\ldots$ is the Euler constant and $\beta$ is
the helium Bethe logarithm \cite{ks} defined as
\ba
\beta &=&
   \frac{\left\langle (\vec{p}_1+\vec{p}_2) (H-E)\ln[2(H-E)]
   (\vec{p}_1+\vec{p}_2)
         \right\rangle}
         {\left\langle (\vec{p}_1+\vec{p}_2) (H-E)
         (\vec{p}_1+\vec{p}_2)
         \right\rangle}\non \\
        &=& 4.370\,039(2).
\label{bl}
\ea
The cited value of $\beta$ has been calculated for the finite mass
of the nucleus. Details of calculations in the limit of no recoil
($M \to \infty$) can be found in \cite{kk}. For convenience of
comparison with earlier results it is worth to write explicitly
the relation to the $Q$-term introduced by Araki and Sucher \cite
{as},
\ba
Q &=& \lim_{\rho \to 0} \left\langle
            \frac{\Theta(r - \rho)}{ 4\pi r^3 }
      + (\ln \rho + \gamma)\delta(\vec{r}) \right\rangle\non \\
  &=& - \frac{1}{2\pi} \left\langle \frac{\ln r + \gamma}{ r^2 }
               i\vec{n} 
                 \vec{p}
               \right\rangle.
\ea

The next, ${\cal O}(\alpha^4)$ correction to the energy is
\cite{heops}:
\ba
&&\delta^{(4)} E = \frac{\alpha^2\delta^{(2)}\!E
        \left\langle c \right\rangle}{2}
        +\alpha^4
   \left\{ - \frac{E^3}{2}
          + \frac{E^2\left\langle c \right\rangle}{4}\right.\non\\
&&      \left. +\; \frac{E}{4}\left\langle 2C_{\rm N} C + c^2
          - \frac{p_1^2 p_2^2}{2} - \pi Z\left[
          \delta(\vec{r}_1)+\delta(\vec{r}_2)
       \right] \right\rangle \right.\non\\[1mm]
&&   +\; \left\langle V_P G V_P \right\rangle
   + \left\langle V_S G V_S \right\rangle\non \\[2mm]
&&   +\; \pi  k_{\rm eN} \langle \delta(\vec{r}_1)
   +\delta(\vec{r}_2) \rangle
   + \pi k_{\rm ee} \langle \delta(\vec{r}) \rangle\non\\[2mm]
&&   + \left\langle - \frac{3 C_1 C_2 C_{\rm N}}{ 4 }
      - \frac{c C_{\rm N} C}{ 2 }
      - \frac{ C_{\rm N} c [\vec{p}_1 \vec{p}_2
      + \vec{n}(\vec{n}\vec{p}_1)\vec{p}_2]
      }{4}
     \right.\non\\
&&    +\; \frac{p_1^2 C_{\rm N} p_2^2}{ 4 }
    + \frac{\vec{p}_1 c^2 \vec{p}_1\!
    + \vec{p}_2 c^2 \vec{p}_2}{ 8 }
    + \frac{(\vec{p}_1\!\times\!\vec{p}_2) c
    (\vec{p}_1\!\times\!\vec{p}_2)}{ 4 }\non \\
&&    - \frac{p_1^2 c\, (\vec{n}\vec{p}_2)^2
    + (\vec{p}_1\vec{n})^2 c\, p_2^2
            - 3 (\vec{p}_1\vec{n})^2 c\,
            (\vec{n}\vec{p}_2)^2
            }{ 8 }\non \\
&&    - \frac{ 2(\vec{n}\vec{p}_2)
    (\vec{E}_1\vec{p}_2)
           + (\vec{n}\vec{E}_1)
           \left[(\vec{n}\vec{p}_2)^2-p_2^2\right]
           }{4}\non \\
&&    +\; r\frac{3 \vec{E}_1\vec{E}_2
    - (\vec{n}\vec{E}_1)(\vec{n}\vec{E}_2)
            - 2 (\vec{E}_1-\vec{E}_2)\vec{e}}{ 8 }
            \non\\
&&            - \frac{ 3 }{ 32 }\frac{ P^2
            -3(\vec{n}\vec{P})^2 }{ r^3 }
            + \frac{ \pi\delta(\vec{r}) }{ 2 }
     \left( \frac{ 9 P^2 }{ 16 } + C_{\rm N} \right)
     \non \\
&&   +\; \frac{\pi Z}{ 4 }\left[ \delta(\vec{r}_1)
          \left( \frac{3p_2^2}{2} - \frac{2Z-1}{r_2} \right)
          + (1\leftrightarrow 2) \right]\non\\
&&   - \frac{ (\vec{E}_1-\vec{E}_2)\vec{e} }{ 32 }
   + \frac{Z^2 }{ 2 } \left[ \frac{1}{r_1^3}
   \left(i\vec{n}_1\vec{p}_1 + Z\right)
      + (1\to 2) \right]\non\\
&&  \left.\left. - \frac{\ln r + \gamma}{2 r^2}i\vec{n}
                 \vec{p}
               + \frac{3}{2r^3} \left( i\vec{n} \vec{p}
               - \frac{1}{2} \right)
          \right\rangle\right\}\non\\
&& = \, 139.60(1)\, \alpha^4. \label{tot6}
\ea
Here we use the following notations: $C = C_{\rm N} + c$, $C_{\rm
N} = C_1 + C_2$, $c=1/r$, $C_{1,2} = -Z/r_{1,2}$, $\vec{e} =
\vec{n}/r^2$ and $\vec{E}_{1,2} = -Z\vec{n}_{1,2}/r_{1,2}^2$. The
terms $\left\langle V_P G V_P \right\rangle$ and $\left\langle V_S
G V_S \right\rangle$ in (\ref{tot6}), where $G$ is the reduced
Green function of the Schr\"odinger equation,
$(H-E)G(\vec{r}_1,\vec{r}_2|\vec{r}_1',
\vec{r}_2')=\psi(\vec{r}_1,\vec{r}_2) \psi(\vec{r}_1',\vec{r}_2')
- \delta(\vec{r}_1-\vec{r}_1')\delta(\vec{r}_2-\vec{r}_2')$,
represent the effects of virtual transitions into triplet $P$ and
singlet $S$ excited states, respectively (see \cite{heops} for
details). Perturbations which induce those transitions are
\begin{equation}
V_P =  \frac{ \vec{s}_1-\vec{s}_2 }{ 4 }
             \left( \frac{ Z
              \vec{l}_1 }{ r_1^3 }
            - \frac{ Z \vec{l}_2 }{ r_2^3 }
            + \frac{ \vec{r}\times\vec{P} }{ r^3 } \right),
\end{equation}
where $\vec{l}_{1,2} = \vec{r}_{1,2} \times \vec{p}_{1,2}$, and
\ba
&& V_S =  E \frac{ C_{\rm N} + 2c }{2}
            + \frac{\{p_1^2 + c, p_2^2 + c\}}{8}
            - \frac{C_{\rm N}c}{2} - \frac{3c^2}{4}\non\\
&&     + \frac{\vec{p}_1\left( C_{\rm N} - c \right)\vec{p}_1
        \! + (1\!\to\! 2)}{4}
        - \frac{ \vec{p}_1 c \vec{p}_2 +
        (\vec{p}_1\vec{n})c(\vec{n}\vec{p}_2) }{ 2 }.\!
\ea
The contact terms enter into Eq.(\ref{tot6}) with the coefficients
\ba
k_{\rm eN} &=& \frac{Z^3}{2} + \frac{427 Z^2}{96}
         - \frac{10 Z}{27}
         - \frac{9 Z \zeta(3)}{4\pi^2}
         - \frac{2179 Z}{648\pi^2}
         \non\\
&& + \frac{3Z - 4Z^2}{2}\ln 2 \to 16.3557, \label{ken}
\ea
\ba
k_{\rm ee} &=& - \ln\alpha
         + \frac{3285}{216}
         - \frac{335}{54\pi^2}
         - \frac{29\ln 2}{2}
         + \frac{15 \zeta(3)}{4\pi^2}\non\\
&& \to 10.37657. \label{kee}
\ea
In (\ref{tot6}), all momentum operators standing to the right
(left) of position-dependent operators are assumed to act on the
right (left) wave function.

Average values for a part of the operators entering into
Eq.(\ref{tot6}) can be found in the literature. For the average
values of the new operators we have obtained the following
results:
$$
\left\langle V_P G V_P \right\rangle = -0.392;\qquad \left\langle
V_S G V_S \right\rangle = - 18.48;\non
$$
$$
\left\langle - \frac{ C_{\rm N} c [\vec{p}_1 \vec{p}_2
      + \vec{n}(\vec{n}\vec{p}_1)\vec{p}_2]
      }{4}\right\rangle = 0.811;\non
$$
$$
\left\langle \frac{p_1^2 C_{\rm N} p_2^2}{ 4 } \right\rangle =
-36.983;\; \left\langle \frac{\vec{p}_1 c^2 \vec{p}_1 + \vec{p}_2
c^2 \vec{p}_2}{ 8 } \right\rangle = 1.142;
$$
$$
\left\langle \frac{(\vec{p}_1\times\vec{p}_2) c
    (\vec{p}_1\times\vec{p}_2)}{ 4 } \right\rangle = 1.078;
$$
$$
\left\langle - \frac{p_1^2 c (\vec{n}\vec{p}_2)^2\!
    + (\vec{p}_1\vec{n})^2 c p_2^2
            - 3 (\vec{p}_1\vec{n})^2 c
            (\vec{n}\vec{p}_2)^2
            }{ 8 }\right\rangle = -0.03;
$$
$$
\left\langle - \frac{ 2(\vec{n}\vec{p}_2)
    (\vec{E}_1\vec{p}_2)
           + (\vec{n}\vec{E}_1)
           \left[(\vec{n}\vec{p}_2)^2-p_2^2\right]
           }{4}\right\rangle = -4.749;
$$
$$
\left\langle r\frac{3 \vec{E}_1\vec{E}_2
    - (\vec{n}\vec{E}_1)(\vec{n}\vec{E}_2)
            - 2 (\vec{E}_1-\vec{E}_2)\vec{e}}{ 8 }
            \right\rangle = 1.434;
$$
$$
\left\langle - \frac{ 3 }{ 32 }\frac{ P^2
            -3(\vec{n}\vec{P})^2 }{ r^3 } \right\rangle =
            -0.353;\;
            \left\langle \frac{ 9\pi\delta(\vec{r}) P^2}{ 32 }
      \right\rangle = 1.325;
$$
$$
\left\langle \frac{ \pi\delta(\vec{r}) C_{\rm N}}{ 2 }
      \right\rangle = -2.508;\;
      \left\langle - \frac{ (\vec{E}_1-\vec{E}_2)\vec{e} }{ 32 }
\right\rangle = 0.409;
$$
$$
      \left\langle \frac{3\pi\delta(\vec{r}_1) p_2^2}{ 2 }
          + (1\leftrightarrow 2) \right\rangle = 18.421;
$$
$$
\left\langle \frac{3\pi\delta(\vec{r}_1) C_2}{ 2 }
          + (1\leftrightarrow 2) \right\rangle = -25.066;
$$
$$
\left\langle \frac{3}{2r^3} \left( i\vec{n} \vec{p}
               - \frac{1}{2} \right)
          \right\rangle = -0.958;
$$
$$
\left\langle \frac{1}{r_1^3}
   \left(i\vec{n}_1\vec{p}_1 + Z\right)
      + (1\to 2) \right\rangle = 4.706.
$$

Finally, the enhanced by $\ln^2 \alpha$ (and hence presumably the
leading) part of the ${\cal O}(\alpha^5)$ correction is
\cite{logs}:
\begin{equation}
\delta^{(5)} E\! =\! -4 Z^3 \alpha^5\ln^2(Z\alpha)
               \left\langle \delta(\vec{r}_1)\! +\!
               \delta(\vec{r}_2) \right\rangle
               \approx 2\,070\, \alpha^5.
\end{equation}

Numerical results for all the contributions to the helium
ionization potential are collected in the Table. Appropriate
expression for the ground state energy of the helium ion is
\ba
&& E_{{\rm He}^+} =
        - \frac{Z^2}{2} \frac{M}{M + 1}
        - \frac{Z^4 \alpha^2}{8} \frac{M(M^2+3M+5)}{(M+1)^3} \non\\
&&- \frac{4Z^4 \alpha^3}{3\pi}
           \left(1-\frac{3}{M}\right)
           \left[ 2\ln (Z\alpha) + \beta_{\rm H}
           - \frac{19}{30} \right] \non\\
&&  - \frac{2Z^5 \alpha^3}{3\pi M}
           \left( \ln (Z\alpha)
               + 4\beta_{\rm H} - 7\ln 2
               - \frac{31}{3} \right) \non\\
&&      + Z^3 \alpha^4 \left( k_{\rm eN} - \frac{9Z^3}{16} \right)
        - \frac{4 Z^6 \alpha^5}{\pi} \ln^2 (Z\alpha)
        + \frac{2 Z^4 r_{\rm N}^2}{3}.
\ea
Here $\beta_{\rm H} = 2.9841285557655\ldots$ is the Bethe
logarithm for the hydrogen ground state and $r_{\rm
N}=3.162(2)\cdot 10^{-5}$ is the nucleus charge radius $R_{\rm N}$
expressed in the atomic units.

A comparison of our result (\ref{result}) and the most recent
experimental values,
\begin{equation}
\nu_{\rm exp}^{1S-2P}(1^1 S) = 5\;945\;204\;238\,(45)\,{\rm MHz},
\label{eikema}
\end{equation}
and
\begin{equation}
\nu_{\rm exp}^{1S-2S}(1^1 S) = 5\;945\;204\;356\,(48)\,{\rm MHz},
\label{bergeson}
\end{equation}
extracted from the measurements of $1^1 S - 2^1 P$ \cite{eik} and
$1^1 S - 2^1 S$ \cite{berg} intervals, respectively, shows that
the theoretical value (\ref{result}) agrees well with the former
(\ref{eikema}) and is within $2\sigma$ from the latter
(\ref{bergeson}) if the theoretical and experimental uncertainties
are added linearly. Efforts in both theoretical and experimental
directions are desirable in order to further clarify the
situation.

\begin{table}
\caption{Contributions to the total ionization potential of the
helium ground state. Uncertainty in the nonrelativistic value is
due to uncertainty in the nucleus mass.} \label{tab}
\begin{center}
\begin{minipage}{0.48\textwidth}
\begin{tabular}{l r@{\hspace{0.3pt}\hspace{0.3pt}}l}
 & \multicolumn{2}{r}{$\delta \nu_{\rm th}(1^1S)$, MHz} \\
\hline \vrule height11pt width 0pt Nonrelativistic approximation
            & $5\;945\;262\;288$.& 62(4)  \\
$\alpha^2$          & $-16\;800$.& 338(4)  \\
$\alpha^3$          & $-40\;486$.& 375(50)  \\
$\alpha^4$          &     $-834$.& 9(2)   \\
$\alpha^5\ln^2 \alpha$ (and higher) &    $84$\hspace{2pt} & (42)   \\
Finite charge radius &     $-29$.& 55(4)  \\
\hline \vrule height11pt width 0pt total       &
$5\;945\;204\;221$\hspace{2pt} & (42) \\
\end{tabular}
\end{minipage}
\end{center}
\end{table}


\subsection*{Acknowledgments}

V.~K. was supported by the National Science Foundation through a
grant for the Institute for Theoretical Atomic and Molecular
Physics at Harvard University and Smithsonian Astrophysical
Observatory, which is gratefully acknowledged. A.~Y. is grateful
to V.~F. Dmitriev, L.~F. Hailo, the late V.~B. Telitsyn, V.~V.
Vecheslavov and O.~V. Zhirov for advices concerning numerical
calculations and to K. Melnikov for comments on the manuscript.
A.Y. was partially supported by the Russian Ministry of Higher
Education and by the Russian Foundation for Basic Research under
grant number 00-02-17646.

\end{document}